\documentstyle[11pt,aaspp4]{article}
\input epsf.sty
\input{psfig}
\newcommand\scr{{\scriptstyle}}
\newcommand\Msun{M_{\sun}}
\newcommand\tp{{\raise 0.3ex\hbox{\fiverm +}}}
\newcommand\th{{\thinspace}}
\newcommand\sep{{\par \noindent \hangindent=15pt \hangafter=1}}
\newcommand\etal{{{\it et al}.\ }}
\newcommand\eg{{{\it e.g.}\ }}

\newcommand\pn{{\par\noindent}}
\newcommand\ie{{{\it i.e}.}}

\newcommand\vecomega{{\rlap{$\omega$}{\hskip 0.1ex\hbox{$\omega$}}}}
\newcommand\vecOmega{{\rlap{$\Omega$}{\hskip 0.1ex\hbox{$\Omega$}}}}
\newcommand\vecSigma{{\rlap{$\Sigma$}{\hskip 0.1ex\hbox{$\Sigma$}}}}
\newcommand\eql {\  = \ }

\newcommand\Omegadot{{\hbox{$\dot \Omega$}}}

\newcommand\edot{{\hbox{$\dot e$}}}

\newcommand\adot{{\hbox{$\dot a$}}}

\newcommand\hdot{{\hbox{$\dot h$}}}

\newcommand\addot{{\hbox{$\dot d$}}}
\newcommand\omegadot{{\hbox{$\dot \omega$}}}

\newcommand\bfOmegadot{{\hbox{$\dot \vecOmega$}}}
\newcommand\dbfdot{{\hbox{$\dot {\bf d}$}}}

\newcommand\dbfddot{{\hbox{$\ddot {\bf d}$}}}
\newcommand\Edot{{\hbox{$\dot {\cal E}$}}}
\newcommand\hbfdot{{\hbox{$\dot {\bf h}$}}}
\newcommand\ebfdot{{\hbox{$\dot {\bf e}$}}}

\newcommand\etadot{{\hbox{$\dot {\eta}$}}}
\newcommand\psidot{{\hbox{$\dot {\psi}$}}}
\newcommand\h{{{\bf h}}}
\newcommand\e{{{\bf e}}}
\newcommand\q{{{\bf q}}}
\newcommand\f{{{\bf f}}}
\newcommand\dbf{{{\bf d}}}
\newcommand\vbf{{{\bf v}}}
\newcommand\r{{{\bf r}}}

\newcommand\halfline{{\null \vskip 7 truept}}
\newcommand\w{{\ \ \ \ \ }}

\newcommand\tle{{\raise 0.3ex\hbox{$\scr {\th \le \th }$}}}
\newcommand\tge{{\raise 0.3ex\hbox{$\scr {\th \ge \th }$}}}
\newcommand\tl{{\raise 0.3ex\hbox{$\scr {\th < \th }$}}}
\newcommand\ts{{\raise 0.3ex\hbox{$\scr {\th \sim \th }$}}}
\newcommand\Chi{{\raise 0.4ex\hbox{$\chi$}}}
\newcommand\Chidot{{\hbox{$\dot {\Chi}$}}}
\newcommand\EEdot{\hbox{$\dot E$}}
\newcommand\Hbf{{\bf H}}
\newcommand\E{{\bf E}}
\newcommand\Q{{\bf Q}}
\newcommand\ftfe{\th{\bf f}_{\rm TF}}
\newcommand\fqde{\th{\bf f}_{\rm QD}}
\newcommand\fqd{${\bf f}_{\rm QD}$}
\newcommand\ftf{${\bf f}_{\rm TF}$}
\newcommand\Opar{\Omega_{\parallel}}
\newcommand\Oper{\Omega_{\perp}}

\newcommand\tehat{{\overline{\it e}}}
\newcommand\tqhat{{\overline{\it q}}}
\newcommand\hhat{{\overline{\bf h}}}
\newcommand\ehat{{{\overline{\bf e}}}}
\newcommand\qhat{{{\overline{\bf q}}}}
\newcommand\Hhat{{{\overline{\bf H}}}}
\newcommand\Ehat{{{\overline{\bf E}}}}
\newcommand\Qhat{{{\overline{\bf Q}}}}
\newcommand\rbar{{{r_*}}}

\begin{document}

\title{The Equilibrium-Tide Model for Tidal Friction}
\author{Peter P. Eggleton, Ludmila G. Kiseleva}
\affil{ Institute of Astronomy, Cambridge CB3 0HA, UK}
\and
\author {Piet Hut}
\affil{ Institute of Advanced Studies, Princeton}
\centerline{e-mail: ppe@ast.cam.ac.uk}
\begin{abstract}
 We derive from first principles equations governing (a) the 
quadrupole tensor of a star distorted by both rotation and the presence of a 
companion in a possibly eccentric orbit, (b) a functional form for the
dissipative force of tidal friction, based on the concept that the rate of 
energy loss from a time-dependent tide should be a positive-definite function
of the rate of change of the quadrupole tensor as seen in the frame which 
rotates with the star, and (c) the equations governing the rates of change of
the magnitude and direction of the stellar rotation, and the orbital period and
eccentricity, based on the concept of the Laplace-Runge-Lenz vector. Our analysis leads
relatively simply to a closed set of equations, valid for arbitrary inclination
of the stellar spin to the orbit. The results are equivalent to classical
results based on the rather less clear principle that the tidal bulge lags 
behind the line of centres by some time determined by the rate of dissipation;
our analysis gives the effective lag time as a function of the dissipation rate
and the quadrupole moment.
\par We discuss briefly some possible applications of the formulation.

\end{abstract}

\keywords{stars: binary, triple}

\section{Introduction}
We present a partly novel derivation of some basic results, and some new
results, for the `equilibrium tide' model of tidal friction (Hut 1981). We 
derive the force and couple
on a binary orbit that arises from dissipation of the equilibrium tide, starting
only from the principles that (a) the rate of dissipation of energy should be a 
positive definite function of the rate of change of the tide, as viewed in a
frame which rotates with the star, and (b) the total
angular momentum is conserved. This model leads to a force and couple which
are similar to those derived classically on the assumption that the tidal bulge
lags the line of centres by some small time that is determined by the timescale
of dissipation. However in our model the lag time is determined to be 
proportional to the quadrupole moment, rather than assumed to be independent of
it. There is a clear relation between the rate of dissipation and the lag 
time. We also determine the relation between the hypothesised dissipation constant and the turbulent  viscosity within the star. However there remains
the more difficult task of determining the turbulent velocity field.
\par There are two principle formalisms for discussing the effect of tidal
dissipation on an orbit: the `equilibrium tide' model (Darwin 1880, Alexander
1973, Hut 1981, Hut 1992) and the `dynamical tide' model (Fabian, Pringle \& Rees 1975, Zahn 1975, Press \& Teukolsky 1977, McMillan, McDermott \& Taam 1987,
Goldreich \& Nicholson 1989, Lai 1996, Kumar \& Goodman 1997, Kumar \& Quataert 1997, Papaloizou \& Savonije 1997, Savonije et al. 1983, 1984, 1995, 1997, 
Mardling 1995). In the former the star is perceived as 
being perturbed to a shape closely approximating the shape it would have in 
equilibrium with the (time-dependent) gravitational-centrifugal potential of 
the system. Dissipation by viscosity of the internal motion caused by the 
time-variation of the tide causes the orbit to tend towards circularity, and 
causes the spin to be brought into parallelism with the orbit, until the tide 
ceases to be time-dependent in the limit that the orbit is circular and the 
spin parallel to the orbit (Hut 1980). In the latter formalism, the star is perceived as 
an oscillator with a number of normal modes that are excited by the 
time-dependent potential as the companion passes by at periastron.  In a close 
encounter from an initially parabolic (or highly eccentric) orbit, energy is 
extracted from the orbital motion and locked up in the oscillation of the star's normal modes. Each mode is damped at some rate, the higher-order modes being 
damped more rapidly. If the modes are all damped during the long apastron phase, then the process can repeat itself at the next periastron, and thus represents a long-term drain of orbital energy intothermal energy. Since the  angular
momentum is constant while energy is dissipated, we expect the orbit to become
more circular. If the modes damp slowly, they may still be active at the next 
periastron, and because the frequencies of oscillation are not necessarily 
commensurate with the frequency of the orbit, each periastron passsage may be 
different from the previous one; some oscillations may be in phase and excited 
resonantly, while others may be out of phase. It is possible for the orbit to be decircularised instead of circularised, and in some circumstances it may vary 
chaotically. Even for circular orbits, dynamical tides can continue to operate, 
as long as there still are deviations from synchronous rotation (Zahn 1975, Goldreich \& Nicholson 1989). 
\par These two models are not necessarily in competition. Rather, it is likely
that the first prevails in cases where an orbit is only moderately eccentric,
and reasonably wide in the sense that both stars are small compared with their
periastron separation. The latter may prevail in circumstances (`tidal capture')
where two stars approach initially on a parabolic or slightly hyperbolic orbit,
and interact briefly and dramatically near periastron. In the former case it is 
likely that only the lowest order (quadrupolar) mode will be significantly
excited, as in the case of tides on Earth. We concentrate here on the
equilibrium-tide model, which we develop using a mathematical-physical 
treatment which we believe to be unusually clear and simple.
\par In $\S$ 2, we derive an approximation to the quadrupole tensor of the
mass distribution in a star of mass $m_1$ distorted by the combination of (a) 
uniform rotation $\vecOmega$, and (b) the presence of a companion of mass $m_2$ 
at position \dbf(t)  relative to the centre of $*1$. $\vecOmega$ is not 
necessarily perpendicular to \dbf, and neither is $\dbfdot$. We deal only
with the lowest, \ie\ quadrupole, mode because it is much the simplest; we
can operate entirely within the linear regime, and extract results that
are exact to first order. Including higher modes involves ultimately working 
in the non-linear regime, and one gains only a modest increase of accuracy, 
if any, at the expense of considerable extra complexity. In $\S$ 3, we show
that if the mechanical energy of the motion (orbital energy plus rotational 
energy) is dissipated at a rate which is proportional to the square of the
rate of change of the quadrupole tensor, then a force (and associated couple)
is introduced into the system which is equivalent to the assumption that the
tidal bulge lags behind the line-of-centres by some time which depends on the
rate of dissipation. In $\S$ 4, we use the concept of the Laplace-Runge-Lenz
(LRL) vector
to determine relatively easily the effect on the binary orbit of an arbitrary 
perturbing force. In $\S$ 5 we apply this analysis to the dissipative force
obtained in $\S$ 3, in combination with the non-dissipative force that is
already present simply because of the quadrupolar distortion of $\S$ 2. 
The standard equations are recovered relatively simply in the case that 
$\vecOmega$ is parallel to the orbital angular momentum; and we obtain the
complete equations for the case that they are not parallel. In $\S$ 6
we derive the tidal velocity field and its rate of dissipation by turbulent
viscosity. This leads to a specific value for the dissipation constant of
$\S$ 3. In $\S$ 7
we determine briefly the meridional circulation current (Eddington 1925, Vogt
1925, Sweet 1950, Mestel 1965) that is
implicit in the distortion determined in $\S$ 2. However this circulation
current must be taken with a pinch of salt since it shows several singularities
in the outer layers of hot stars. In $\S$ 8 we discuss uncertainties in the
model, and mention briefly possible applications of the model to (a) the 
circularisation of a pulsar orbit around a normal star, and (b) the effect on a 
close binary orbit of the presence of a distant third star in a hierarchical 
triple system.
\section{The Equilibrium Tide}
 Consider a binary consisting of an extended star ($*1)$, which  rotates 
with constant uniform angular velocity $\vecOmega$, and a point-mass companion 
($*2$) at a distance $\dbf(t)$ from it. In a frame which rotates with $*1$, and
has origin at the CM of $*1$, the equation of motion of the fluid of $*1$ is
$${D\vbf\over Dt}+2\vecOmega\wedge\vbf\eql  - {1\over\rho}\nabla p-\nabla\phi\w,\w 
\phi\eql \phi_1-{1\over 2}\vert\vecOmega\wedge\r\vert^2-{Gm_2\over\vert\r+\dbf\vert}
-{Gm_2\ \r.\dbf\over d^3}\w ,\eqno(1a,b)$$
where \r\ is a vector from the CM of the extended star to a general point in it, and $D/Dt$ is the derivative following the motion. The potential $\phi$ is the 
combination of the potential of self-gravity $\phi_1$, where
$$\nabla^2 \phi_1\eql  4\pi G\rho\w ,\eqno(2)$$
along with centrifugal force, and with the tidal potential of the companion. 
The last term in Eq. (1b) allows for the fact that the frame centred at the
CM of $*1$ is accelerated relative to an inertial frame; it cancels the
first-order term in $r/d$ from the second-last term. To lowest order in $r/d$, 
$$\phi\ \approx\ \phi_1-{1\over 3}\Omega^2r^2+{1\over 3}\Omega^2r^2 
P_2(\cos\theta)-{Gm_2r^2\over d^3}P_2(\cos\theta^{\prime})\w ,\eqno(3)$$
where $\theta$ is the polar angle measured from $\vecOmega$, and
$\theta^{\prime}$ is the polar angle measured from \dbf. 
\par If the spin and orbit are not aligned, and/or the orbit is elliptical, the 
star cannot be entirely static, \ie\  $\vbf\not=0$, but we assume that \vbf\ is 
small so that at some level of approximation pressure balances rotational and
gravitational forces:
$$\nabla p \  = \  -\rho \nabla \phi\w . \eqno(4)$$
Hence $p,\rho$ are constant on equipotentials.  We can then define variables 
$V, m, \rbar$ -- volume, mass and `volume radius' respectively -- which are 
also constant on equipotentials. $V$ is the volume contained within an 
equipotential and $m$ the mass, and they relate to $\rbar$  by
$${4\pi \over 3}\rbar^3\eql V(\phi)\w,\w {dm\over d\rbar}\eql 4\pi\rbar^2\rho(\phi)
\w . \eqno(5)$$
We can see that, since the distance between adjacent equipotentials along a 
normal is $\delta l = \delta\phi/\vert\nabla\phi\vert$,
$${dV\over d\phi}\  = \  \int{\delta l\over \delta\phi} \th d\Sigma\  = \  \int 
{d\Sigma\over \vert\nabla\phi\vert}\ \  ,\eqno(6)$$
where $d\Sigma$ is an element of area.
\par The constancy of $p,\rho$ on equipotentials, and hence of $T$ also at
least in the case of chemically homogeneous stars, means that in the heat
equation we have to introduce a velocity field, the `circulation current',
to maintain thermal equilibrium ($\S$ 7). We assume this field is 
sufficiently small that we can still neglect $D\vbf/Dt,\vecOmega\wedge\vbf$ in
Eq. 1a. The theory developed in $\S\S$ 2 -- 6 does not require any
knowledge of the circulation velocity, only supposing that it is small
so that the approximation of $p,\rho$ constant on equipotentials is valid.

\par The following subsections (i) -- (viii) develop the following points as
briefly as possible:
\sep (i) a definition of the quadrupolar-distortion parameter $\alpha(r)$
\sep (ii) the quadrupole moment as an integral of $\alpha$
\sep (iii) the surface boundary condition for $\alpha$, in the case of rotation
but no companion
\sep (iv) the second-order DE satisfied by $\alpha$ in the interior
\sep (v) analogous to (iii), but for companion and no rotation
\sep (vi) an alternative form of the surface boundary condition (Schwarzschild
1958)
\sep (vii) an illustrative but too crude approximation for $\alpha$
\sep (viii) the potential force $\fqde$ (in addition to point-mass gravity) between bodies when one has a quadrupolar distortion due to both rotation and
a companion.

\pn (i) We consider only the lowest-order, quadrupolar, distortion. An equipotential surface $r=r(\rbar,\theta)$ will be approximated by
$$r\ \approx\ \rbar\{1-\alpha(\rbar) P_2(\cos\theta)\}\w ,\w{\rm and\ conversely}\w\rbar\ \approx\ 
r\{1+\alpha(r) P_2(\cos\theta)\}\w,\eqno(7a,b)$$
where $\theta$ is the angle from the axis of symmetry. Since $\alpha$ is first
order, it can be thought of as a function of either $r$ or $\rbar$. 

\pn (ii) The contribution to the quadrupole moment $q$ from mass between
equipotential surfaces $\phi$ and $\phi+d\phi$ is
$$ {dq\over d\phi} \eql  \int \rho r^2 P_2(\cos\theta){d\Sigma\over\vert\nabla\phi \vert} \eql  \rho(\phi)\int{2\pi r^2\sin\theta d\theta
\over \vert\nabla\phi\vert}\  r^2P_2(\cos\theta)\ \  , \eqno(8)$$
where we use Eq. (6) to estimate the volume element. Now, $\phi$ is a function 
of $\rbar$ only, and $\rbar$ is a function of $r,\theta$ given by Eq. (7b), so that
$$\vert\nabla\phi\vert\eql {d\phi(\rbar)\over d\rbar}\vert\nabla\rbar\vert\ \approx
\ {d\phi(\rbar)\over d\rbar}\left[1+{d\th r\alpha(r)\over dr} P_2\right]\ \approx\ {d\phi 
(\rbar)\over d\rbar}\left[1+{d\th \rbar\alpha(\rbar)\over d\rbar} P_2\right]\w.\eqno(9)$$
Hence
$$ {dq\over d\rbar}\ \approx\ \rho(\rbar)\int\left[1-{d\th\rbar\alpha(\rbar) \over 
d\rbar} P_2)\right]\rbar^4(1-4\alpha P_2)P_2\th 2\pi\sin\theta d\theta\eql  
-{4\pi\over5}\rho \rbar^4 (\left[4\alpha +{d\th\rbar\alpha\over d\rbar}\right], \eqno(10)$$
and so
$$q\eql -{1\over 5}\int_0^{m_1}(5\alpha+\rbar\alpha^{\prime})\rbar^2dm
\w.\eqno(11)$$

\pn (iii) The  self-gravity term $\phi_1$ of Eq. (3), {\it outside} the star, is just the
sum of the monopole and quadrupole terms, and so the centrifugal-gravitational 
potential distribution outside the star is
$$ -\phi({\bf r}) \  \approx \  {Gm_1\over r}+{G q^{\rm rot}\over r^3}
P_2(\cos\theta)+{1\over 3}\Omega^2r^2-{1\over 3}\Omega^2r^2 P_2(\cos\theta)
\w . \eqno(12)$$
If $r_1$ is the radius of $*1$, this means that on the stellar surface 
($\rbar\eql r_1)$, using Eq. (7a), 
$$-\phi\ \approx\ {Gm_1\over r_1}+{1\over 3}\Omega^2 r_1^2+\left(\alpha_1 
{Gm_1\over r_1}+{G q^{\rm rot}\over r_1^3}-{1\over 3}\Omega^2r_1^2\right) 
P_2(\cos\theta)\eqno(13)$$
where $\alpha_1\equiv\alpha(r_1)$. This must be independent of $\theta$, so that
$$ \alpha_1\eql r_1^3 \left({\Omega^2 \over 3Gm_1}-{q^{\rm rot}\over m_1 r_1^5} 
\right)\w . \eqno(14)$$
Using Eq. (11), this leads to
$$\alpha_1\eql {\Omega^2r_1^3\over 3Gm_1}\ {1\over 1-Q}\ ,
\w q^{\rm rot}\eql -{\Omega^2 r_1^5\over 3G}\ {Q\over 1-Q}\ ,\w {\rm with}\w 
Q\equiv{1\over 5}{\int r^2 dm
(5\alpha+r\alpha^{\prime})\over m_1r_1^2\alpha(r_1)}\w. \eqno(15a,b,c)$$
Once again, since $\alpha$ is first order we do not need to distinguish between
$\alpha(\rbar)$ and $\alpha(r)$ in the integral.
\pn (iv) Evidently to determine $Q$ we now have to determine the functional 
form of $\alpha(r)$. From Eq. (7b) throughout the interior, rather than 
just at the surface, we can obtain $\nabla^2\rbar$ and $\vert\nabla\rbar
\vert^2$ to first order in $\alpha$ and its derivatives, and since $\phi$ 
as well as $\rho$ is a function of $\rbar$ only we find that
$$\nabla^2\phi\eql \phi^{\prime\prime}\vert\nabla\rbar\vert^2+\phi^{\prime}\nabla^2
\rbar\eql \phi^{\prime\prime}+{2\over\rbar}\phi^{\prime}+\left[\phi^{\prime}(r
\alpha^{\prime\prime}+4\alpha^{\prime}-{2\alpha\over r})
+2\phi^{\prime\prime}(r\alpha^{\prime}+\alpha)\right]\phi^{\prime}P_2$$
$$=4\pi G\rho(\rbar)-2\Omega^2\w. \eqno(16)$$
Outside the square brackets a prime means a derivative w.r.t. $\rbar$, but 
inside it can mean either $r$ or $\rbar$, since the expression is already 
first order. For Eq. (16) to be true for all $\theta$ we require firstly that 
$$\phi^{\prime\prime}+{2\over\rbar}\phi^{\prime}\eql 4\pi G\rho-2\Omega^2\w,
\w{\rm or\ to\ zero\ order}\w 
\phi^{\prime}\ \approx\ {Gm(r)\over r^2}\w,\eqno(17a,b)$$
and secondly, from the $P_2$ coefficient, we obtain after some rearrangement
$$ \alpha^{\prime\prime}-{6\alpha\over r^2}+{2rm^{\prime}\over m}\left( 
{\alpha^{\prime}\over r}+{\alpha\over r^2}\right)\eql 0\w, \eqno(18)$$
where we only need the zero-order approximation (17b) to eliminate $\phi$,
since $\alpha$ is already first-order. We can therefore determine $\alpha$ 
by first solving the usual stellar structure equations to obtain $m(r), 
m^{\prime}(r)$, and then integrating Eq. (18) with these functions, subject 
to boundary conditions that $\alpha$ and $\alpha^{\prime}$ are finite at the 
centre. These boundary conditions determine $\alpha$ up to a multiplicative 
constant. For at $r=0$, we have $rm^{\prime}/m=3$. Hence from Eq. (18) 
$\alpha\ts B+C/r^5$, and so the central boundary conditions require $C=0$,
leaving only an arbitrary multiplicative constant. Consequently $Q$ is determined unambiguously, 
since the expression (15c) for $Q$ is independent of a constant factor in 
$\alpha$. For a polytrope of index $n$ in the range 0 -- 4.95, we find that 
$Q$ as given by Eqs. (15c) and (18) can be approximated by the interpolation formula
$$Q\ \approx\ {3\over 5}\th\left(1-{n\over 5}\right)^{2.215}\th 
e^{0.0245n-0.096n^2-0.0084n^3}\ \ \pm 1.5\%\ {\rm r.m.s.}.\eqno(19)$$
\pn (v) For a star distorted by the gravitational field of a companion, and 
{\it not} rotating, the calculation is the same as in (iii) except that, by 
reference to Equn (3), in the potential (13) we must replace $\Omega^2/3$ by $-Gm_2/d^3$ inside the parentheses; outside the parentheses we put $\Omega=0$. 
Also, of course, the angle on which $P_2$ depends is measured now from the line 
of centres, rather than from the axis of rotation.  By analogy with Eq. (15),
$$ \alpha_1\eql -\th{m_2r_1^3\over m_1d^3}{1\over 1-Q}\w ,\w q^{\rm comp}\eql  
{m_2 r_1^5\over d^3} {Q\over 1-Q}\   \ \  , \eqno(20a,b)$$
with the same structure constant $Q$ as before. 
\pn (vi) Schwarzschild (1958) obtained, by a slightly different route, a result 
which in our notation is
$$q^{\rm comp}\eql {m_2r_1^5\over d^3}\th\left({3\alpha-r\alpha^{\prime}\over
2\alpha+r\alpha^{\prime}}\right)_{r=r_1}\w.\eqno(21)$$
Although superficially very different from Eq. (20b) with $Q$ given by Eq. (15c), it is
in fact the same by virtue of the fact that
$${d\over dr}\th mr^2(3\alpha-r\alpha^{\prime})\eql r^2(5\alpha+r\alpha^{\prime})
m^{\prime}\w,\eqno(22)$$
as can be verified by using Eq. (18) to eliminate $\alpha^{\prime\prime}$
from the LHS of Eq. (22). Thus
$$Q\eql \left({3\alpha-r\alpha^{\prime}\over 5\alpha}\right)_{r=r_1}\eql {1\over 5m_1
r_1^2\alpha_1}\th\int(5\alpha+r\alpha^{\prime})r^2dm\w.\eqno(23)$$
\pn (vii) We can obtain a crude but illustrative approximation to $\alpha$ by
assuming that (a) the mass is concentrated at the centre, and (b) the 
quadrupole is concentrated at the surface, where the distortion is greatest. 
Then the potential {\it inside} the star $(\rbar\tle r_1)$ -- {\it cf.} 
Eq. (12) for {\it outside} -- is given by
$$-\phi({\bf r})\ \approx\ {Gm_1\over r}+{G q^{\rm rot}r^2 P_2(\cos\theta)
\over r_1^5}+{1\over 3}\Omega^2r^2-{1\over 3}\Omega^2r^2 P_2(\cos\theta)
\w . \eqno(24)$$
The condition that $\phi$ is constant on any interior potential 
$\rbar\eql $const. implies that
$$-\phi(\rbar)\ \approx\ {Gm_1\over\rbar}+{1\over 3}\Omega^2\rbar^2+\left(
\alpha {Gm_1\over\rbar}+{G q^{\rm rot}\rbar^2\over r_1^5}-{1\over 3}
\Omega^2\rbar^2 \right)P_2(\cos\theta) \eqno(25)$$
is independent of $\theta$ and hence that
$$\alpha\eql \rbar^3\left({\Omega^2\over 3Gm_1}-{q^{\rm rot}\over m_1r_1^5}
\right) \ \approx\ r^3\left({\Omega^2\over 3Gm_1}-{q^{\rm rot}\over 
m_1r_1^5} \right) \w . \eqno(26)$$
Eq. (14) is just Eq. (26) evaluated at $\rbar\eql r_1$. Putting 
$\alpha\propto r^3$ in Eq. (15c), we obtain an approximation for $Q$ which we
can evaluate numerically for polytropes and check against Eq. (19):
$$ Q\ \approx\ {8\over 5}{\int r^5dm\over r_1^5m_1} \ \approx\ 
{3\over 5}\th\left(1-{n\over 5}\right)^{2.205}\th 
e^{-0.437n+0.066n^2-0.023n^3}\ \ \pm 3\%\ {\rm r.m.s.}\w . \eqno(27)$$
The approximation $\alpha\propto r^3$ that we have obtained can be seen to 
satisfy Eq. (18) in the limit of a centrally-condensed star, where $m^{\prime}
= 0$ everywhere except at a central point. The full solution in this
case is $\alpha\eql Br^3+C/r^2$. We cannot assume that $C$ is negligible 
despite the fact that it leads to singularity at the origin, since 
$m^{\prime}$ is also singular at the origin. However in practice, $C$ does 
become negligible in polytropes as $n\to 5$.
\par Note that the approximation $\alpha\propto r^3$ of Eq. (26) gives zero if we use the differential form of Eq. (23), as against Eq. (27) if we use the integral form of Eq. (23). The differential form requires us to obtain a more accurate solution for $\alpha$ than does the integral form, as is not unusual.
\par  In the extreme that the star is of constant density, $rm^{\prime}/m\eql 3$ 
throughout and Eq. (18) gives $\alpha=B+C/r^5$. This means that $\alpha$ is constant, 
since the solution must be non-singular at the origin. This is just the 
well-known case of `liquid' stars. For such stars (Maclaurin ellipsoids, Chandrasekhar 1969) all the 
equipotentials are similar ellipsoids with eccentricity $e$ where
$$ {\Omega^2r_1^3\over Gm_1}\eql {3\sqrt{1-e^2}\over 2e^3}\{(3-2e^2)\sin^{-1}e
-3e\sqrt{1-e^2}\}\ \approx\ {2\over 5}e^2\w. \eqno(28)$$
The quadrupole moment of a uniform ellipsoid is
$$q^{\rm rot}=-{1\over 5}m_1r_1^2e^2(1-e^2)^{1/3}\ \approx\ -{\Omega^2r_1^5\over 2G}\w.\eqno(29)$$
This `agrees' with Eqs. (15b) and (27) in the case that $\rho =$const. and hence $Q=3/5$. 
The agreement is providential, however, since we ought to use Eq. (15c) with $\alpha =$const., whereas Eq. (27) assumed $\alpha\propto r^3$; but both expressions give the same answer if $\rho =$const. This agreement, providential or not, coupled with the genuine agreement for centrally condensed stars ($n\to 5)$, appears to suggest that 
approximation (27) might be good enough over the whole range of models from uniform density to centrally condensed.  However if, for polytropes, we compare 
the $Q$ of Eq. (19), obtained from the direct numerical solution of Eqs. (18) 
and (15c), with $Q$ obtained from Eq. (27), using the fitting formulae as given, we find that the disagreement can approach $\ts 50\%$
for the important range of $n\ts$ 1.5 -- 3. We therefore recommend against using
approximations (26) and (27), convenient as they are for illustrative purposes.
\pn (viii) Putting together Eqs. (15) and (24), and recalling that the symmetry 
axes of the two quadrupoles are $\vecOmega$ 
and $\dbf$, we see that the quadrupole tensor of a star in a binary can be written
$$q_{ij}\eql -{A\over 6G} \th(3\Omega_i\Omega_j-\Omega^2\delta_{ij})+{Am_2\over 
2 d^5}\th(3 d_id_j-d^2 \delta_{ij})\ ,\w A\eql {r_1^5Q\over 1-Q} \ ,\eqno(30a,b)$$
where $Q$ might be given either by Eqs. (15c) and (18), or by approximation
(27).
\par The quadrupolar distortion of Eq. (25) produces an acceleration, 
in addition to the gravity of two point masses, which we call $\f_{\rm QD}$. 
This non-dissipative term is
$$\mu\fqde\eql -\nabla \delta\Phi,\w-\delta\Phi\eql \th{Gm_2 d_id_jq_{ij}\over d^5}
\eql  -{m_2A(\vecOmega.\dbf)^2\over 2d^5}+{m_2A\Omega^2 \over 6d^3}+ 
{Gm_2^2 A\over d^6}\ \  ,\eqno(31)$$
and so the orbital equation is
$$\dbfddot\eql   -\th{Gm\dbf\over d^3}\th +\th \f_{\rm QD} \w, \eqno(32)$$
where
$$\f_{\rm QD}\eql  {m_2A\over \mu}\left[{5(\vecOmega.\dbf)^2\dbf\over 2d^7} -{\Omega^2\dbf\over 2d^5}-{\vecOmega.\dbf \th\vecOmega\over d^5} 
-\th{6Gm_2 \dbf\over  d^8}\right] \w .\eqno(33) $$
\section{Dissipation of the Equilibrium Tide}
\par We continue to suppose that $*1$ is extended while 
$*2$ is a point mass.  In the frame which rotates with $*1$, the quadrupolar 
tide will in general be time-dependent: $*1$ will be continually changing 
its shape. If the star is not perfectly elastic we expect a loss of total 
mechanical energy, but no loss of total angular momentum for the system as a whole.
\par In the previous $\S$ we had an acceleration \fqd, Eq. (33), which is 
derivable from a potential $\delta\Phi$, Eq. (31). Under prevailing 
circumstances this conserves total energy, (including rotational energy), as we 
show shortly. But if there is in addition a slow dissipation of energy by tidal 
friction, we will have an extra acceleration \ftf, say. Writing
$$ \dbfddot \  = \  -{Gm \dbf \over d^3}-{1\over\mu}\nabla\delta\Phi + \ftfe\w,  
\eqno(34)$$
we can see that total angular momentum
$$\Hbf\ \equiv\ \mu\dbf\wedge\dbfdot+I\vecOmega \eqno(35)$$
is conserved if the couple on $*1$ is given by 
$$I\bfOmegadot\eql \dbf\wedge(\nabla\delta\Phi-\mu\ftfe)\w.\eqno(36)$$
The total energy,
$$E\eql {\mu\over 2}\dbfdot .\dbfdot - {Gm\mu\over d} 
+\delta\Phi(d,\vecOmega.\dbf) + {1\over 2} I\Omega^2\w,\eqno(37)$$
may change not only because \dbf\  varies but also because
\vecOmega\  can vary, for
instance as a result of precession. We find however that $\EEdot\eql 0$ (in
the absence of tidal friction) provided that $\delta\Phi$ depends only on 
$d$ and $\vecOmega.\dbf$:
$$\EEdot\eql \mu\dbfdot.\{\dbfddot+{Gm \dbf \over d^3}+{1\over\mu}\nabla\delta\Phi\}
+\bfOmegadot.(\dbf\th\th\delta\Phi^{\prime}+I\vecOmega)$$
$$\hskip 0.3truein\eql \mu\dbfdot.\ftfe+{\dbf\over I}\wedge(\nabla\delta\Phi-
\mu\ftfe).(\dbf\th\th\delta\Phi^{\prime}+I\vecOmega)\w,$$
$$\eql \mu(\dbfdot-\vecOmega\wedge\dbf).\ftfe+\nabla\delta\Phi\th .
\th\vecOmega\wedge\dbf\hskip 0.7truein \eqno(38)$$
where we have used Eqs. (34) and (36) for \dbfddot\  and \bfOmegadot. The 
quantity $\delta\Phi^{\prime}$ means the partial derivative of $\delta\Phi$ 
with respect to \vecOmega.\dbf. For $\delta\Phi$ a function only of the quantities $d$ 
and $\vecOmega.\dbf$, we see that $\nabla\delta\Phi$ is entirely in the plane of 
\dbf\ and \vecOmega, and hence that the last term in Eq. (38) vanishes:
$$\EEdot\eql \mu{\partial\dbf\over\partial t} .\ftfe\w,\w{\rm since}
\w\dbfdot\eql {\partial\dbf\over\partial t}+\vecOmega\wedge\dbf\w,\eqno(39a,b)$$
where $\partial/\partial t$ is a derivative in the frame rotating with
$*1$ and $\dbfdot$ is the derivative in an inertial frame.
\par When there is dissipation, probably the simplest assumption we can make is 
that the rate of loss of energy is some positive-definite function of the rate 
of change (as seen in the frame rotating with the star) of $*1$'s shape, \eg of 
its quadrupole tensor since this determines its shape to lowest order. We therefore write
$$ \EEdot\eql  -\sigma\th{\partial q_{ij}\over \partial t}
\th{\partial q_{ij}\over \partial t}\w ,\eqno(40)$$
where $ q_{ij}$ is given by Eq. (30) and $\sigma$ is a dissipation constant 
intrinsic to $*1$ (dimensions $m^{-1}l^{-2}t^{-1}$). In the frame that rotates 
with $*1$, \vecOmega\ is approximately constant
while \dbf\ varies at rate $\partial\dbf/\partial t$, so that 
$${\partial q_{ij}\over \partial t}\eql {\partial \over \partial t}\th{m_2A\over 
2d^5} (3d_id_j-d^2\delta_{ij})\hskip 1truein\eqno(41)$$
$$\hskip 1truein\eql {3m_2A\over 2d^5}\left[ d_i{\partial d_j\over \partial t}+d_j{\partial 
d_i\over \partial t}+d{\partial d\over \partial t}\delta_{ij}-{5\over d}
{\partial d\over \partial t} d_id_j\right]\ \  .\eqno(42)$$
Squaring this, after some manipulation we obtain
$$\EEdot\eql -{9\sigma m_2^2A^2\over 2d^8}\left[2\left({\partial d\over 
\partial t}\right)^2+\left({\partial\dbf\over\partial t}\right)^2\right]\eql 
-{9\sigma m_2^2A^2\over 2d^{10}}{\partial{\bf d}\over\partial t}. 
\left[2\th\dbf\th\dbf\th.{\partial{\bf d}\over \partial t}+d^2{\partial{\bf d}\over 
\partial t}\right]\ \ . \eqno(43)$$
Comparing this with Eq. (39a) we see that a consistent expression for the 
acceleration $\f_{\rm TF}$ due to tidal friction is
$$ \f_{\rm TF}\eql -{9\sigma m_2^2A^2\over 2\mu d^{10}} \left[2\th\dbf\th\dbf 
\th. {\partial{\bf d} \over \partial t}+d^2{\partial{\bf d}\over \partial t} 
\right] \eqno(44)$$
$$\hskip 0.5truein\eql -{9\sigma m_2^2A^2\over 2\mu d^{10}} \left[3\th\dbf\th\dbf\th. 
\th\dbfdot+ (\h-\vecOmega d^2)\wedge\dbf \right]\ \  .\eqno(45)$$
\par The mathematical form of this expression for the perturbing force due to 
tidal friction is exactly the same as that obtained by the more traditional 
model in which it is assumed that the tide, which would be directed along the line of 
centres if it were able to adjust instantaneously to the potential, is offset 
by some small lag time that is proportional to the timescale of dissipation. It
is not clear that this lag time should be independent of orbital phase, but such
an assumption is justified on the basis of our own model.
\section{The Effect of a Perturbing Force on the Orbit}\par Using the relative position vector ${\bf d} \equiv {\bf d}_1-{\bf d}_2$, 
and mass $m \equiv m_1 +m_2$, the relative motion of a binary subject to (i)
Newtonian point-mass gravity, and (ii) an additional acceleration {\bf f} (the 
perturbing force per unit reduced mass $\mu \equiv m_1 m_2/m$) is given by
$$ \dbfddot \  = \  -{Gm \dbf \over d^3} + \f \ \  . \eqno(46)$$
A remarkably straightforward way to analyse the effect of a general (small)
force \f\  on the otherwise Keplerian orbital solution is by way of the
LRL vector (Heggie, personal communication, 1996). This has many 
applications apart from the present one (Eggleton 2000, in preparation), and
so we describe it in a slightly more general way than is strictly necessary
for present purposes.
\par Define ${\cal E}$ (Keplerian energy, per unit reduced mass), 
{\bf h} (angular momentum, similarly) and {\bf e} (LRL vector) by
$${\cal E}\ \equiv\  {\scr 1 \over \scr 2}\dbfdot .\dbfdot - {Gm \over d}\  , 
\w {\bf h}\ \equiv \ {\bf d}\wedge \dbfdot \  , \w \ Gm\e \ \equiv\  \dbfdot 
\wedge \h - {Gm\dbf \over d} \ . \eqno(47a-c) $$
Note that ${\cal E}$ is not the {\it total} energy, if $\f\not= 0$, but only the
part that is kinetic plus Newtonian point-mass energy.
We can see, after some manipulation in the case of {\bf e}, that
$$ \Edot \  = \  \dbfdot . \f \  , \w \hbfdot \ =\  \dbf \wedge \f\  , \w Gm\ 
\ebfdot\  = \  \f \wedge \h + \dbfdot \wedge (\dbf\wedge \f) \  . \eqno(48a-c)$$
Hence ${\cal E}$, {\bf h} and {\bf e} are all constants of the motion if 
{\bf f}=0. Using a standard parametrisation of the Keplerian orbit, for example 
either of the two parametrisations (51) - (55) below, we find that the vector 
{\bf e}  defined in Eq. (47c) is in the direction of periastron, and has 
magnitude equal to
the eccentricity (thus justifying belatedly the choice of name {\bf e}). In 
this section we show how $\Edot$, $\hbfdot$ and $\ebfdot$ can be estimated by 
averaging the LHS of Eqs (48a) - (48c) over a Keplerian orbit, supposing \f\ is small.
\par Even if \f\  is not small, auxiliary variables $a$ (semimajor axis), $b$ 
(semiminor axis), $l$ (semilatus rectum), $\omega$ (mean angular velocity) and $p$ 
(period) can be {\it defined} in terms of ${\cal E}, \h, \e$ in the usual way:
$$a \eql  -{Gm \over 2{\cal E}}\  , \w b \eql  a\sqrt{1-e^2} \  , \w l\eql  a(1-e^2) 
\  , \w \omega \eql   {h\over ab}\eql  {2\pi \over p} \  . \eqno(49a-d)$$
For general \f , and not just \f\  $\approx$ 0, four standard relations can be 
shown to be satisfied identically:
$$2h^2{\cal E}+G^2m^2(1-e^2)\eql 0\w,\w\e\th.\th\h\eql 0\w,\w h^2=Gml \w,\w \omega^2=
Gm/a^3\w.\eqno(50a-d)$$
Thus even when \h, $l,\omega$ and $a$ are continuously changing because $\f\not=
0$,  the orbit can be perceived as always `instantaneously Keplerian'. 
For example, the instaneous period and semimajor axis always satisfy Kepler's
third law, Eq. (50d).
\par If \f \ is small, we can estimate its effect on ${\cal E}, \h, \e \ $ by 
averaging over time the RHS's of Eqs (48) in a Keplerian orbit. This is done 
most easily by writing the Keplerian orbit in Cartesians (origin at focus, 
\e\ in the $1$-direction, ${\bf q}\ \equiv \h\wedge\e$ in the 2-direction, 
\h\  in the $3$-direction) using one or other of the following parametric forms:
$$\dbf  \eql  \w {l \over 1+e \cos \theta}(\cos \theta , \sin \theta, 0) \eql  (a\cos \phi - ae, b\sin \phi, 0)\hskip 0.5truein\eqno(51) $$
$$\dbfdot  \eql  \w {\omega ab\over l}\th (-\sin \theta, \cos \theta+e, 0)\eql  {\omega \over 1-e\cos \phi}\th (-a\sin \phi , b\cos \phi, 0)\eqno(52) $$
$$\omega dt \eql  \w {l^2 \over ab}\th {d\theta \over (1+e\cos \theta)^2} \hskip 0.5truein\eql  (1-e\cos \phi) d\phi\hskip 1.2truein \eqno(53) $$
$$ d\eql  \w {l \over (1+e\cos \theta)}\hskip 0.5truein\eql  a(1-e\cos \phi) \hskip 1truein \eqno(54)$$
$$\addot\eql \w{\omega abe\over l} \sin \theta\hskip 0.7truein\eql  {\omega ae
\sin\phi\over1-e\cos\phi}\hskip 1.1truein  \eqno(55)$$
Various scalar, vector and tensor functions of \th \dbf \  can be averaged over 
time using these parametrisations, $\theta$ being more useful if the function 
contains a substantial negative $(\tle -2$) power of $d$ and $\phi$ otherwise 
$( \tge -1)$. Since \e, \q$\ \equiv\h\wedge\e$ and \h\  are not unit vectors, we use bars to indicate the corresponding unit vectors; then for example
$d_i=d\tehat_i\cos\theta+d\tqhat_i\sin\theta$. Some specific averages, expressed in terms of polynomials $I_{n,l}(e), \ n,l\tge 0,$ defined below, are as follows:
$$<{1\over d^{n+2}}>\eql {1\over p}\int_0^p{dt\over d^{n+2}}\eql {1\over abl^n} \int_0 ^{2\pi}{d\theta\over(1+e\cos\theta)^n}\eql {1\over abl^n}\th I_{n,0}\hskip 1.2truein\eqno(56)$$
$$<d^{n-1}>\eql a^{n-1}I_{n,0} \hskip 4.0truein   \eqno(57)$$
$$<{\addot_i\addot_i\over d^{n+2}}> \eql {\omega^2ab\over l^{n+2}}\{(1+e^2)I_{n,0}
+2eI_{n,1}\}\hskip 2.7truein\eqno(58)$$
$$<d^n\addot_i\addot_i> \eql  \omega^2a^{n+2}\{I_{n,0}-eI_{n,1}\}\hskip 3.2truein\eqno(59)$$
$$<{\addot^2 \over d^{n+2}}> \eql  {\omega^2ab e^2 \over l^{n+2}} \th 
\{I_{n,0}-I_{n,2}\} \hskip 3.2truein\eqno(60)$$
$$<{d_i\over d^{n+3}}>\eql 
{1\over p}\left[\tehat_i\int_0^p{\cos\theta\th dt\over d^{n+2}}+\tqhat_i
\int_0^p{\sin\theta\th dt\over d^{n+2}}\right]\eql 
{1\over abl^n}\th I_{n,1}\tehat_i\hskip 1.3truein\eqno(61)$$
$$<d^{n-1}d_i>\eql -a^n\{I_{n,1}+eI_{n,0}\}\tehat_i \hskip 3.3truein\eqno(62)$$
$$<{\addot_i\over d^{n+2}}> \eql {\omega\over l^{n+1}}\{eI_{n,0} 
+I_{n,1}\} \tqhat_i \hskip 3.2truein\eqno(63)$$
$$<{\addot d_i \over d^{n+3}}>\eql  {\omega e\over l^{n+1}}\th\{I_{n,0}
-I_{n,2}\}\tqhat_i \hskip 3.25truein\eqno(64)$$
$$<{d_id_j\over d^{n+4}}>
\eql {1\over p}\left[\tehat_i\tehat_j\int_0^p{\cos^2\theta\th dt\over d^{n+2}} +(\tehat_i\tqhat_j+\tqhat_i\tehat_j)\int_0^p{\cos\theta\sin\theta\th dt\over d^{n+2}}
+\tqhat_i\tqhat_j\int_0^p{\sin^2\theta\th dt\over d^{n+2}}
\right]$$
$$\eql {1\over abl^n}\{I_{n,2}\tehat_i\tehat_j
+(I_{n,0}-I_{n,2}) \tqhat_i\tqhat_j\}\eqno(65)$$
$$<{d_id_jd_k\over d^{n+5}}>\eql  {1\over abl^n}\{(I_{n,1}-I_{n,3}) 
(\tehat_i\tqhat_j\tqhat_k+\tqhat_i\tehat_j\tqhat_k+\tqhat_i\tqhat_j
\tehat_k) +I_{n,3}\tehat_i\tehat_j\tehat_k\}\hskip 0.7truein \eqno(66)$$
$$<{d_id_j\addot_k\over d^{n+4}}>\eql {\omega\over l^{n+1}} \{(I_{n,3}-
I_{n,1})(\tehat_i\tqhat_j\tehat_k+\tqhat_i\tehat_j\tehat_k)+(I_{n,1}+eI_{n,0})
\tqhat_i\tqhat_j\tqhat_k\hskip 1.1truein $$
$$\hskip 1.2truein +(I_{n,3}+eI_{n,2})(\tehat_i\tehat_j\tqhat_k
-\tqhat_i\tqhat_j\tqhat_k)\}\w.\eqno(67)$$
The polynomials $I_{n,m}$, all positive, are defined by
$$ I_{n,m}(e)\equiv\int_0^{2\pi}\th(1+e\cos\theta)^n\th\cos^m\theta\th{d\theta 
\over 2\pi}\ .\eqno(68)$$
They are easily evaluated from
$$I_{0,2m}\eql \int_0^{2\pi} \cos^{2m}\psi\th{d\psi \over 2\pi}\eql {(2m)!\over 
2^{2m}(m!)^2}\w, \w I_{0,2m+1}\eql  0 \ ,\w I_{n+1,m}\eql I_{n,m}+eI_{n,m+1}\w .
\eqno(69)$$
Clearly
$$\int_0^{2\pi}\th(1-e\cos\phi)^n\th\cos^m\phi\th{d\phi\over 2\pi}\eql I_{n,m}(-e) 
\eql (-1)^m  I_{n,m}(e)\ .\eqno(70)$$
In this paper the results that we need all come from the 
$\theta$-parametrisation, but we include the $\phi$-parametrisation 
in Eqs (51) - (55) for completeness.
\section{The Effect of \fqd\ and \ftf\ on the Orbit}
\par We can now use some of Eqs. (56) - (67) to average the RHSs of Eqs. (40a) - (40c) for the particular forces \fqd, \ftf\  of Eqs. (33) and (45). We 
firstly consider the simpler case where the spin is parallel to the orbit, and
then the more general case.
\subsection {Stellar Spin Parallel to Orbit}
\par We first specialise to the case $\vecOmega\parallel\h$, which means that 
$\vecOmega.\dbf=0$. This puts to zero the most complicated term which appears in 
Eq. (33). Averaging $\Edot$ by Eq. (48a), and consequentially $\adot, 
\omegadot$ by Eqs. (49a), (50d), the potential force \fqd\  gives zero. The 
term $\f_{\rm TF}$ of Eq. (45), using the averages (56) and (60) with Eqs. 
(68), (69), gives
$${2\over 3}{\omegadot\over \omega}\eql -{\adot\over a}\eql {\Edot\over {\cal E}}\eql 
{<\dbfdot.\ftfe>\over {\cal E}}\eql \  -{9\sigma m_2^2A^2\over 2\mu{\cal E}} \left 
[3<{\addot^2\over d^8}> +h^2<{1\over d^{10}}>-\Omega h<{1\over d^8}>\right]$$
$$\eql -{9\sigma m_2^2A^2\over 2\mu{\cal E}}{\omega^2ab\over l^8}\left[3e^2(I_{6,0}
- I_{6,2}) +I_{8,0}-{\Omega l^2\over\omega ab}I_{6,0}\right]\hskip 1 truein$$
$$\hskip 0.4truein\eql {1\over t_{\rm TF}}\left[{1+{31\over 2}e^2+{255\over 8}e^4+
{185\over 16}e^6+{25\over 64}e^8\over (1-e^2)^{15/2}}\  -\  {\Omega\over\omega}
{1+{15\over 2}e^2+{45\over 8}e^4+{5\over 16}e^6\over (1-e^2)^6}\right]\ \ ,\eqno(71)$$
where the timescale $t_{\rm TF}$ is defined, using Eqs (25), (49) and (50), by
$$t_{\rm TF}\eql {\mu a^8\over 9\sigma m_2^2A^2}\eql  {1\over 9\sigma}\th {a^8\over 
r_1^{10}}\th {m_1\over m_2m}\th\left({1-Q\over Q}\right)^2\ \  .\eqno(72)$$
\par Similarly we average $\hbfdot$ with Eq. (48b). Once again $\f_{\rm QD}$ 
does not contribute, this time because it is purely radial if $\vecOmega.\dbf=0$;
but \ftf, using Eqs. (56), (68) and (69), gives the orbital average
$$\hbfdot\eql <\dbf\wedge\ftfe>\eql  -{9\sigma m_2^2A^2\over 2\mu}\h\left[<{1\over d^8}>
-{\Omega\over h}<{1\over d^6}>\right]\hskip 1truein$$
$$\eql  -{9\sigma m_2^2A^2\h\over2\mu abl^6}\left[I_{6,0}-{\Omega l^2\over
\omega ab} I_{4,0}\right]\hskip 1.5truein\eqno(73)$$
\ie
$${\hdot\over h}\eql -{1\over 2t_{\rm TF}}\left[{1+{15\over 2}e^2+{45\over 8}e^4+ 
{5\over 16}e^6\over (1-e^2)^{13/2}}-{\Omega\over\omega}{1+3e^2+
{3\over 8}e^4\over (1-e^2)^5}\right]\w.\eqno(74)$$
\par We can also average $\ebfdot$ using Eq. (48c). We consider separately the
terms from \fqd\ and \ftf. Since in the present case ($\vecOmega\parallel\h,
\vecOmega.\dbf=0$) \fqd\ is parallel to \dbf, we obtain from Eq. (33) the orbital average
$$Gm\ebfdot_{\rm QD}\eql <\fqde>\wedge\h\eql {m_2A\Omega^2\over 2\mu}\h\wedge 
<{\dbf\over d^5}>+{6Gm_2^2A\over \mu}\h\wedge<{\dbf\over d^8}>\w,\eqno(75)$$
and so using Eqs (49), (61), (68) and (69),
$$\ebfdot_{\rm QD}\ \equiv\ Z\ \hhat\wedge\e\w,\w
Z\eql {mA\over 2m_1a^5\omega}\left[{\Omega^2\over (1-e^2)^2}+{30m_2\omega^2\over m}
{1+{3\over 2}e^2+{1\over 8}e^4 \over (1-e^2)^5}\right]\ . \eqno(76)$$
The vector \e\  is in the direction of the line of apses, and thus $Z$ is the 
rate of apsidal motion, \ie\ the rate at which the line of apses rotates about 
the direction of \h.
\par The effect of \ftf\ on the average of $\ebfdot$ is to produce 5 terms:
$$\ebfdot_{\rm TF}\ \propto\ 3<{\dbf\addot\over d^9}>\wedge \h+h^2<{\dbf\over 
d^{10}}>-\Omega h<{\dbf\over d^8}>+ <{\dbfdot\over d^8}>\wedge\h\ -<{\dbfdot
\over d^6}>\wedge\vecOmega\ .\eqno(77)$$
By reference to Eqs. (64), (61) and (63), all these terms are ultimately
parallel to \e, \ie\ they circularise the orbit without (further) apsidal 
motion. In this case we can obtain $\edot$ more simply than $\ebfdot$ by 
differentiating Eq. (50a), and using Eqs. (71) and (74):
$${\edot\over e}\eql -{1-e^2\over e^2}\left({\Edot\over 2{\cal E}}+{\hdot\over h} 
\right)\eql -{9\over 2t_{\rm TF}}\left[ {1+{15\over 4}e^2+{15\over 8}e^4+
{5\over 64}e^6\over (1-e^2)^{13/2}}\ -\ {11\Omega\over 18\omega}{1+{3\over 2}e^2 +{1\over 8}e^4\over (1-e^2)^5}\right]\ \  .\eqno(78)$$
\par We can obtain the rate of change of the
intrinsic spin $\Omega$, using the constancy of $\Hbf$ in Eq. (35):
$${\Omegadot\over\Omega}\eql -{\mu h\over I\Omega}\ {\hdot\over h}\eql {\gamma\over 2
t_{\rm TF}}\left[{1+{15\over 2}e^2+{45\over 8}e^4+{5\over 16}e^6\over 
(1-e^2)^{13/2}}-{\Omega\over\omega}{1+3e^2+{3\over 8}e^4\over(1-e^2)^5} 
\right]\ \  , \ \  \gamma\ \equiv\ {\mu h\over I\Omega}\ \ ,\eqno(79)$$
where $\hdot/h$ comes directly from Eq. (74). The factor $\gamma$, the ratio of
orbital to spin angular momentum, is usually large. As a result, the ratio
$\Omega/\omega$ tends rapidly to a `pseudo-synchronous' value (Hut 1981) 
which depends only on $e$ and is obtained by equating to zero the contents 
of the brackets in Eq. (79). Then, on a slower time scale, $e\to 0$ and 
$\Omega\to\omega$.
\subsection{ The Non-Aligned Case}
\par For the general case where $\vecOmega\wedge\h\not= 0$ there are rather more
terms, but the problem is quite tractable. Eqs. (48) give $\Edot$, $\ebfdot$  
and $\hbfdot$ in terms of the perturbing force; we average these
over a Keplerian orbit, using the forces \fqd\ and \ftf\  of Eqs. (33), (45).
The recipes of Eqs. (56) - (70) allow us to express the results in the form
$$\Edot/{\cal E}=U\eqno(80)$$
$$\ebfdot/e\eql -V\ehat+Z\qhat-Y\hhat\eqno(81)$$
$$\hbfdot/h\eql Y\ehat-X\qhat-W\hhat\ \ ,\eqno(82)$$
where the coefficients $U, ..., Z$ may be somewhat complicated as functions of 
$e$, {\it via} the integrals $I_{n,m}(e)$, but are otherwise simple functions of $a,b,l,\omega, \sigma, ...$.  The coefficients ($X,Y,Z)$ are in fact a vector, 
the angular velocity of the \e, \q, \h\ frame relative to an inertial frame: in 
$\S$ 5.1, $X=Y=0$, and $Z$ gives the term for apsidal motion in Eq. (76). The evaluation of $U, ..., Z$ is in our experience most easily done by splitting 
$\vecOmega$ into a component $\vecOmega_{\parallel}\equiv\vecOmega.\hhat\th\hhat$ 
parallel to \h, and the corresponding perpendicular component $\vecOmega_{\perp}$. 
Then we split each of the forces \fqd\ and \ftf\ into two parts. One part, say 
$\delta$\fqd\ or $\delta$\ftf,  is defined as 
$$\delta{\bf f}_{\rm QD}\ \equiv\ {m_2A\over\mu}\left[{5(\vecOmega_{\perp}.\dbf)^2
\over 2d^7}-{\vecOmega_{\perp} .\dbf\vecOmega\over d^5}\right]\w,\w
\delta{\bf f}_{\rm TF}\ \equiv\  {9\sigma m_2^2 A^2 \over 2\mu}\th
{\vecOmega_{\perp}\wedge\dbf\over d^8}\w.\eqno(83)$$
These vanish when $\vecOmega_{\perp}\eql 0$. The other part is the remainder. This 
remainder gives the same values for $U\eql \Edot/{\cal E},\ V\eql \edot/e,$ and 
$W\eql \hdot/h$ as Eqs. (71), (74), (78), except that $\Omega$ is to be replaced by $\Opar$. It also gives the same apsidal motion $Z$ as Eq. (76), {\it without} 
replacing $\Omega$ by $\Opar$. The extra terms ($\delta$\fqd, $\delta$\ftf) 
brought into $\hbfdot$ by non-alignment are all in the plane of \e, \q, and 
contribute only to rotation.  
\par It is helpful to use Euler angles $\eta,\Chi,\psi$ say to determine 
the orientations of $\ehat, \qhat, \hhat$ relative to an inertial frame,
say $\Ehat, \Qhat, \Hhat$, a suitable choice for \Hbf\ being the total angular
momentum vector of Eq. (35):
$$\Hbf\eql \mu\h+I\vecOmega\w.\eqno(84)$$
\E\ and \Q\  are arbitrary, provided they make a right-handed orthogonal set 
with \Hbf. The transformation from $\Ehat, \Qhat, \Hhat$ to $\ehat, \qhat, \hhat$
is the product of three successive simple rotations: by $\Chi$ about $\Hhat$, by
$\eta$ about (new) $\Ehat$, and by $\psi$ about (newer still) $\Hhat$, which
now coincides with $\hhat$. This gives
$$\ehat\eql (\cos\Chi\cos\psi-\sin\Chi\sin\psi\cos\eta,\sin\Chi\cos\psi+\cos\Chi
\sin\psi\cos\eta,\sin\eta\sin\psi)\eqno(85)$$
$$\qhat\eql (-\cos\Chi\sin\psi-\sin\Chi\cos\psi\cos\eta,-\sin\Chi\sin\psi+\cos\Chi
\cos\psi\cos\eta,\sin\eta\cos\psi)\eqno(86)$$
$$\hhat\eql (\sin\eta\sin\Chi,-\sin\eta\cos\Chi,\cos\eta)\w, \eqno(87)$$
where we take the 1,2,3 -- axes in the directions of $\Ehat, \Qhat, \Hhat$. 
By differentiating Eqs. (85) -- (87), it is straightforward to show that 
the angular velocity ($X,Y,Z$) of Eqs. (81), (82) relates to $\etadot,
\Chidot,\psidot$ by
$$X\eql \etadot\cos\psi+\Chidot\sin\psi\sin\eta\eqno(88)$$
$$Y\eql -\etadot\sin\psi+\Chidot\cos\psi\sin\eta\eqno(89)$$
$$Z\eql \psidot+\Chidot\cos\eta\w.\eqno(90)$$
From Eqs. (84) - (87), the components of $\vecOmega$ in the directions
of $\e, \q$ and $\h$ are given by
$$\Omega_e\eql  \vecOmega.\ehat\eql {1\over I}
\th(\Hbf-\mu\h).\ehat\eql\Omega_0\sin\eta\sin\psi\eql\Oper\sin\psi\w,\eqno(91)$$
$$\Omega_q\eql \vecOmega.\qhat\eql \Omega_0\sin\eta\cos\psi\eql \Oper\cos\psi\w,\eqno(92)$$
$$\Opar=\vecOmega.\hhat\eql \Omega_0\cos\eta-\mu h/I\eqno(93)$$
\pn where $\Omega_0$, $\Oper$ have values
$$I\Omega_0\ \equiv\ H\w,\w \Oper\eql \Omega_0\sin\eta\w.\eqno(94)$$
\par To obtain $\hbfdot$ and $\ebfdot$ from Eqs. (48b,c) and (45), we use most 
of the averages (56) -- (67); we have to average quantities which range from 
scalars to tensors of rank 3.  After substantial manipulation of the 
orbit-averaged expressions, which includes a pleasantly surprising amount
of simplification, we obtain
$$\etadot\eql X\cos\psi-Y\sin\psi\eql -{\Oper\over 4\omega t_{\rm TF}}\th 
{1+3e^2+{3\over 8}e^4+{3\over 2}e^2(1+{1\over 6}e^2)\cos2\psi\over (1-e^2)^5}\w,
\eqno(95)$$
$$\Chidot\eql {X\sin\psi+Y\cos\psi\over \sin\eta}\eql -{Am_2\Omega_0\Opar\over 
2\mu\omega a^5(1-e^2)^2}-{3\Omega_0\over 8\omega t_{\rm TF}}\th 
{e^2(1+{1\over 6}e^2)\sin2\psi\over (1-e^2)^5}\w,\eqno(96)$$
and
$$\psidot+\Chidot\cos\eta\eql Z\eql {Am_2(\Opar^2-{1\over 2}\Oper^2)\over 
2\mu\omega a^5(1-e^2)^2}+{15Gm_2^2A\over 
\mu\omega a^8}\th{1+{3\over 2}e^2+{1\over 8}e^4\over (1-e^2)^5}\eqno(97)$$
We see that in the absence of tidal friction ($\sigma=0=1/t_{\rm TF}$) we have 
steady precession, with $\eta$ = const. and
$$\Chidot\eql -{Am_2\Omega_0\Opar\over 2\mu\omega a^5(1-e^2)^2}\eql {\rm const.}
\eqno(98)$$
We also see that Eq. (97) for apsidal motion is remarkably little different
from Eq. (76) derived for the case of spin parallel to orbit. We emphasise that
Eqs. (95) - (97) do not involve any approximation about the smallness of the
inclination $\eta$, or the eccentricity $e$.
\par Our complete set of equations for $\hdot, \edot, \etadot, \Chidot$ and
$\psidot$ are then Eqs. (74) and (78) -- with $\Omega$ replaced by $\Opar$ -- 
and (95) to (97). The ancillary variables $a, \omega, \Opar, \Oper$ and $\Omega_0$  
are given in terms of $h, e, \eta, H$ by Eqs. (49), (50) and (93), (94). $A$ and $t_{\rm TF}$ are given by Eqs. (25) and (72). We suppose that $m_1, m_2, r_1, Q, \sigma, H, I$ are all given constants. 
\par The quantity $\Omega_0$ defined in 
Eq. (94) is usually rather large: the angular velocity that the star would 
have if all the orbital angular momentum were converted into spin. Consequently 
$\eta$, the angle between the orbital angular momentum and the total angular 
momentum, is usually rather small. The inclination $i$ of the rotation axis of 
the star to the orbital axis is given by
$$\Oper\eql \Omega\sin i\eql \Omega_0\sin\eta\w.\eqno(99)$$
\section{The Tidal Velocity Field and the Dissipation Constant}
\par In the above derivation the dissipation constant $\sigma$ was assumed to
be given. However it is presumably determined by the internal physics of $*1$.
The most important ingredients are likely to be (a) viscosity, with turbulent
viscosity expected to be far more significant than molecular viscosity, at 
least in the regions that are turbulent, and (b) thermal diffusion. We shall 
assume here that turbulent viscosity is always the dominant dissipative agent, 
even in stars which are of early spectral type and normally assumed to have
non-convective envelopes. Partly this is because even hot stars  have
at least one and usually two surface convection zones, driven by helium
ionisation and/or bumps in the opacity. These zones may be small in mass but
have considerable convective velocities, which presumably spill over into
adjacent regions; they are also positioned close to the surface where they can
contribute best to tidal dissipation. 
\par A further reason for expecting 
turbulence in the surface layers of upper MS stars is that they are generally
rapidly rotating, and the rotationally-driven circulation currents, even though
small in the interior, can become significant in the layers near the photosphere. In fact, a somewhat literal interpretation of rotationally-driven 
circulation ($\S$ 7 below) gives a velocity field which is singular 
wherever the entropy
gradient is zero (\ie\ at the upper and lower edge of any convection zone) as
well as at the surface where $\rho\eql 0$. Such singularities are unphysical, and
are presumably rendered finite by the turbulent viscosity that is driven by
the large shear in these layers. We expect that the upshot will be an outer
envelope that contains significant turbulent viscosity, and consequently it
seems reasonable to assume that the main agent of dissipation is this
turbulent viscosity. Molecular viscosity is normally negligible in comparison,
and so too is thermal diffusion. We return to this discussion in $\S$ 8.
\par Note that in early-type stars, turbulent friction in the convective core 
may contribute to turbulent viscosity.  This contribution is not expected to be 
very efficient (Zahn 1966). For recent discussions, see Zahn (1989), Goldman \& 
Mazeh (1994), and Goodman (1997).
\par In addition to the circulation velocity field ($\S$ 7) there is, in
the case of eccentric orbits and/or non-corotating stars, a time-dependent tidal velocity field driven by the time-dependent character of the distortion. This 
field can be determined by using (a) the continuity equation
$${\partial\rho\over\partial t}+\nabla.\rho\vbf \eql 0\ ,  \eqno(100)$$
and (b) the constancy of $\rho$ on equipotential surfaces, \ie\  the fact that
$$\rho\eql \rho(\rbar)\w,\w \rbar\eql r+r\alpha P_2(\cos\theta)\ .  \eqno(101a,b)$$
Working in the frame which rotates with $*1$, we need only be concerned with 
the companion-induced distortion, so that $\alpha_1$, the surface value of 
the distortion parameter $\alpha$, is given by Eq. (20a). We are entitled to
ignore a second contribution, from rotation, to the asphericity of $\rbar$ in
Eq. (101b), because our analysis below depends only on $\partial\rbar/\partial t$, and the rotational contribution is of course (nearly) constant in the 
frame that rotates with $*1$.
\newcommand\k{{\bf k}}
\newcommand\kdot{{\hbox{$\dot k$}}}
\newcommand\kbfdot{{\hbox{$\dot {\bf k}$}}}
\par Let
$$F\ \equiv\  r^2P_2\eql {3\over 2}(\k.\r)^2-{1\over 2}r^2\w,\w{\rm so\  that}
\w\nabla^2 F\eql 0\w,\w\r.\nabla F\eql 2F\ ,  \eqno(102)$$
where $\k\equiv\dbf/d$. Since \dbf\  is time-varying, both $\alpha$ and \k\  depend 
on $t$, the former because $\alpha\propto 1/d^3$ (Equn 20a). Then with $\rho$ as
a function of $\rbar$ only, and $\rbar$ viewed
as a function of $\r,t$, we obtain
$${\partial\rho\over\partial t}\eql {d\rho\over d\rbar}\th{\partial\rbar\over
\partial t}\eql {d\rho\over d\rbar}\left({\partial\alpha\over\partial t}{F\over r}+
{3\alpha G\over r}\right)\eql {3\alpha\over r}\th{d\rho\over d\rbar}
\left(-{1\over d}{\partial d\over\partial t}\th F+ G\right)\ ,  \eqno(103)$$
where
$$G\ \equiv\ {1\over 3}{\partial F\over\partial t}\eql  \k.\th\r\ {\partial\k\over
\partial t}.\th\r\ \w,\w{\rm so\  that}
\w\nabla^2 G\eql 0\w,\w\r.\nabla G=2G\ .  \eqno(104)$$
$F$ and $G$ are obviously orthogonal harmonic functions of
degree 2. Consider the velocity field given by
$$\vbf\eql {3\alpha_1\over 2}\beta(r)\left({1\over d}{\partial d\over\partial t}\th
\nabla F-\nabla G\right)\ , \w{\rm so\ that}\w \nabla.\rho\vbf\approx{3\alpha_1
\over r}\th{d\th \rho\beta\over dr}\left({1\over d}{\partial d\over\partial t}
\th F- G\right)\ .\eqno(105a,b)$$
Then Eq. (100) is satisfied to first order, provided that
$${d\th\rho\beta\over dr}\eql {\alpha\over\alpha_1}\th{d\rho\over dr}\w,\w\ie\w \beta\eql {1\over\rho\alpha_1}\int_{r_1}^r\th\alpha{d\rho\over dr}\th dr\w.  
\eqno(106)$$
The lower limit in the integral comes from the boundary condition that the
outer surface ($\rho=0$) is a surface which moves with the fluid, so that the 
velocity must be finite there despite the vanishing density. The function 
$\beta$ is determined unambiguously by the structure of the star, via Eq. (18) 
determining $\alpha(r)$, and is well-behaved for polytropic $(0\tl n\tl 5$) 
surfaces as $\rho\to 0$, despite the apparent singularity there. For the special case $n=0$, \ie\  uniform density, we have $\beta\eql \alpha/\alpha_1\eql 1$. Fig. 1
illustrates the behaviour of $\rho,\alpha$ and $\beta$ as functions of $r$ in
four cases: polytropes of index $n=1.5$ and 3, and ZAMS stars of masses 1 and
8 $\Msun$.
\par In suffices,
$$v_i\eql {3\alpha_1\over 2}\th\beta(r)s_{ij}x_j\w {\rm where}\w s_{ij}\ \equiv\ 
{1\over d}{\partial d\over\partial t} \th(3k_ik_j-\delta_{ij})-k_i{\partial k_j
\over\partial t}-{\partial k_i\over\partial t}k_j\ . \eqno(107)$$
The tensor $s_{ij}$ is symmetric and traceless. This allows us to calculate the 
rate of dissipation of energy due to the action of turbulent viscosity on this 
velocity field, and this dissipation in turn determines the strength of tidal 
friction. 
\par The rate-of-strain tensor is now seen to be
$$t_{ij}\ \equiv\ {\partial v_i\over\partial x_j}+{\partial v_j\over\partial 
x_i}\eql {3\alpha_1\over 2}\th \left(2\beta s_{ij}+{\beta^{\prime}\over r}\{s_{ik}
x_kx_j+s_{jk}x_kx_i\}\right)\ .  \eqno(108)$$
We square this and average it over an equipotential (which at this level of
approximation can be taken to be spherical), using the results
$${1\over 4\pi}\int x_ix_j\th d\Omega\eql {r^2\over 3}\th\delta_{ij}\w,\w{1\over 
4\pi}\int x_ix_jx_kx_l\th d\Omega\eql {r^4\over 15}\th(\delta_{ij}\delta_{kl}+
\delta_{ik}\delta_{jl}+\delta_{il}\delta_{jk})\w.\eqno(109)$$
We obtain
$${1\over 4\pi}\int t_{ij}^2 d\Omega\eql 9\alpha_1^2\ s_{ij}^2\ \left(\beta^2+
{2\over 3}r\beta\beta^{\prime}+{7\over 30}r^2\beta^{\prime 2}\right)\ .  
\eqno(110)$$
Now,
$$s_{ij}^2\eql 6\left({1\over d}{\partial d\over\partial t}\right)^2 +2 \left( 
{\partial \k\over\partial t}\right)^2\eql {2\over d^2}\left[2\left({\partial 
d\over \partial t}\right)^2+\left({\partial\dbf\over\partial t}\right)^2\right]
\ ,  \eqno(111)$$
and so the rate of dissipation of mechanical energy is
$$\EEdot\eql -{1\over 2}\int\th\rho wl\th t_{ij}^2\th dV$$
$$\eql -{9\alpha_1^2\over d^2}
\th \left[2\left({\partial d\over \partial t}\right)^2+\left({\partial\dbf\over
\partial t}\right)^2\right]\int_0^{m_1}wl\th\left(\beta^2+{2\over 3}r\beta
\beta^{\prime}+{7\over 30}r^2\beta^{\prime 2}\right)\th dm\ .  \eqno(112)$$
The parameters $w,l$ are the mean velocity and mean free path of turbulent  
eddies. As we emphasised at the start of this Section, $w$ might be due to
shear in the rotationally-driven circulation velocity field, as well as to
convection (of which there is normally some even in stars with supposedly
radiative envelopes). Fig.~1 shows the how the weight factor in parentheses in 
Eq. (112), which we call $\gamma(r)$, varies throughout some polytropic and 
ZAMS model stars.
\begin{figure}
\plotone{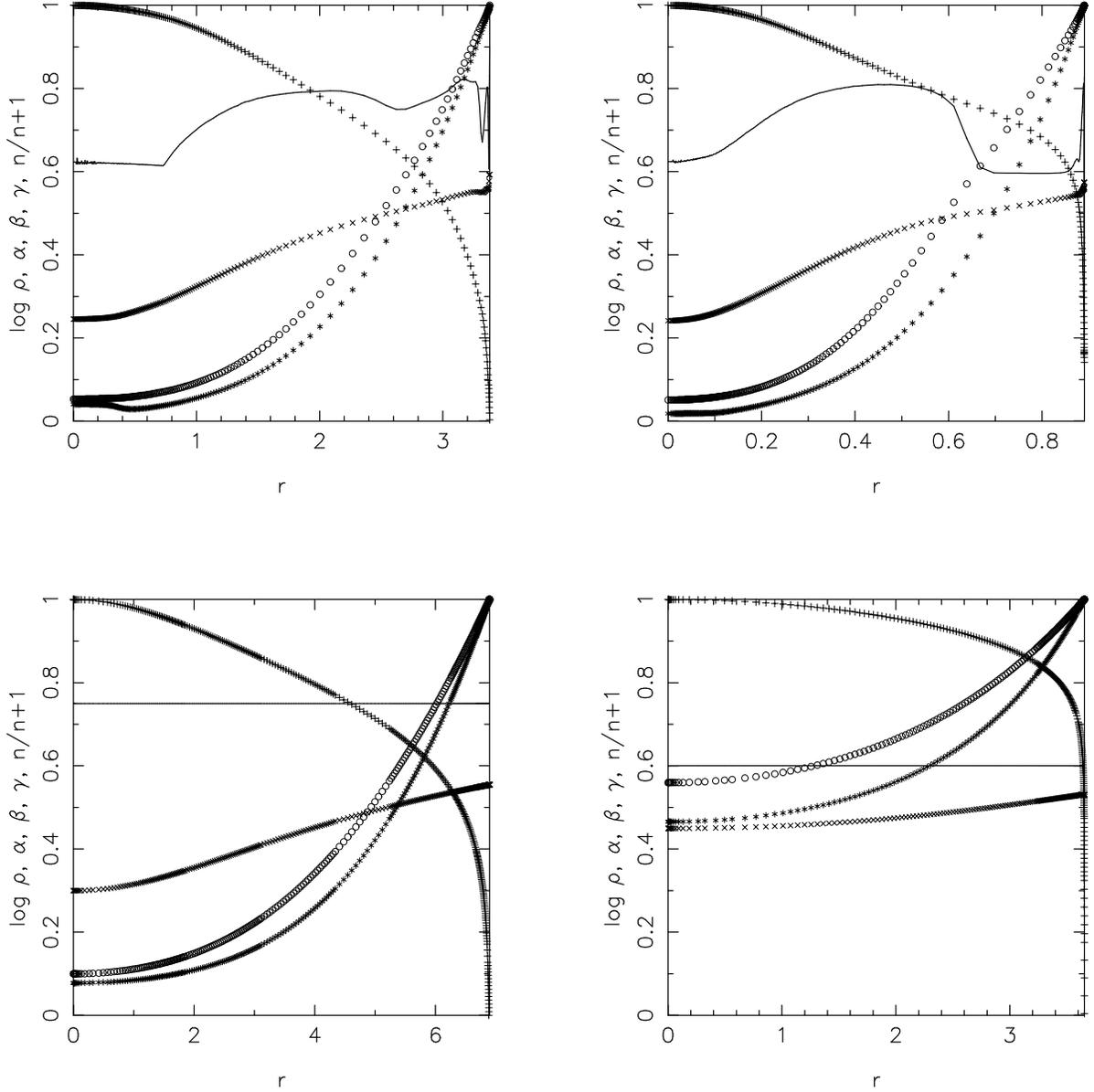}
\caption{The variation of $ 0.1\log(\rho/\rho_c) + 1$ 
(plusses), $\alpha/\alpha_1$ (asterisks) and $\beta$ 
(circles) with radius (arbitrary units); $\rho_c$ is the central 
density. Also shown are $n/n+1$ 
(continuous line), where $n$ is the local effective polytropic 
index, and finally (crosses) $0.1\log\gamma + 0.5$, where 
$\gamma$ is the weight factor in Eq. (113), $i.e.$ 
the contribution which each layer would make to the 
dissipation rate if the viscosity were uniform. (a) $8\Msun$
(b) $1\Msun$ (c) $n=3$ polytrope (d) $
n=1.5$ polytrope. Modest fluctuations in ${n/n+1}$ near the centre
in the two stellar models are due to the fact that this quantity was obtained
by the ratio of differences in a tabular model, which had converged to a
maximum-modulus error of $\sim 10^{-6}$.\label{fig1}}
\end{figure}
\par It is gratifying (but not coincidental) that Eqs. 
(111), (112) contain the same functional dependence on $\dbf(t)$ as Eq. (43),
despite being derived by a very different route. Equating (43) with (112) 
allows us to determine $\sigma$ as 
$$\sigma\eql {2\over m_1^2r_1^4Q^2}\th\int \th wl\th\gamma(r)\th dm\w, \w
\gamma(r)\ \equiv\ \beta^2+{2\over 3}r 
\beta\beta^{\prime}+{7\over 30}r^2\beta^{\prime 2}\w.\eqno(113)$$
This in turn, {\it via} Eq. (72), gives an estimate for the tidal friction
timescale:
$$t_{\rm TF}\eql \left({a\over r_1}\right)^8{m_1^2\over 18mm_2}(1-Q)^2\th
{m_1r_1^2\over\int\th wl\th\gamma(r)\th dm}\w, \eqno(114)$$ 
We defer to $\S$ 8 a brief discussion of the possible significance of this
estimate, but we note here that it does {\it not} depend on any of the following
approximations and assumptions: (i) $\nabla\wedge\vbf\eql 0$, or $\nabla.\vbf\eql 0$,
(ii) $\alpha\propto r^3$, or $\beta\propto r^4$, (iii) $\vecOmega\parallel 
\vecomega$, or $e\ts 0$. It depends only on the approximation that the quadrupole
tide dominates, that it is small enough to validate a linearisation of the
continuity equation (100), and that there can be assumed to be an isotropic
viscosity.
\section{Rotationally Driven Circulation}
\par In the surface layers of rapidly rotating stars, rotationally-driven circulation could well be the second most important source of turbulence, after convection. For a homogeneous (\ie\ zero-age) star, $T$ as well as $p,\rho$ is 
constant on equipotentials, and hence so is $\epsilon$, the rate of release 
of nuclear energy. Consequently we can define a nuclear luminosity $L$ 
interior to an equipotential by
$$ {dL\over dV}\eql \rho\epsilon \ \  .\eqno(115)$$
Let us define
$$K(\phi)\  \equiv \  4\pi Gm-2\Omega^2 V \  =\  \int\nabla\phi .d\vecSigma
\  = \  \int \vert\nabla\phi\vert d\Sigma \eqno(116)$$
and note, from Eqs (1b) and (5), that 
$$\nabla^2\phi={dK\over dV}\ \ .\eqno(117)$$
Using Eq. (6) for the volume element, we see that
hydrostatic equilibrium can be written as
$$-{1\over\rho}{dp\over d\rbar}\eql {d\phi\over d\rbar}\eql {d\phi\over dV} 
{dV\over d\rbar}\eql {4\pi\rbar^2\over\int d\Sigma/\vert\nabla\phi\vert} 
\eql \left({Gm\over\rbar^2}-{2\Omega^2\rbar\over 3} \right){(4\pi\rbar^2)^2\over 
\int\vert\nabla\phi\vert\th d\Sigma\int d\Sigma/ \vert\nabla\phi\vert}\ \  
.\eqno(118)$$
The factor in parentheses simply cancels the first factor in the denominator
to its right, by (116); we write it this way to show that the ratio on the 
right of the parentheses clearly differs from unity in {\it second} order
if $\phi$ differs from spherical in {\it first} order, so that we can write
$$-{1\over\rho}{dp\over d\rbar}\  = \ {d\phi\over d\rbar} \  \approx \  
{Gm\over \rbar^2} - {2\Omega^2 \rbar \over 3}
\ \  .\eqno(119)$$
This is the first integral of Eq. (17a), keeping the first-order term
which was neglected in Eq. (17b).
\par Now consider the energy flux {\bf F}, which in general is a combination of
radiative and convective flux. In spherical symmetry, we usually write this as
$$F \  = \  -{4acT^3 \over 3\kappa \rho} {dT \over dr}\  +\  \rho wT\delta S
\eqno(120)$$
where $w$ is the mean velocity of convection and $T\delta S$ is the mean heat 
excess of an upward-rising eddy. The mixing-length approximations for $w$ and 
$T\delta s$ are
$$w^2 \  \ts \  T\delta S \  \ts \  Tl\left[-{dS\over dr}\right]\  \ts \ 
T\left[{dS\over d\log p}\right] \ \  , \eqno(121)$$
where $l$ is the mixing length, normally estimated by $l \ts -dr/d\log p$, and
where the square brackets have the meaning $\left[X\right]\ \equiv\
{\rm max}(X,0)$. In a non-spherical situation the generalisation of the 
radiative term in the energy flux is obvious; and of the many possible 
generalisations for the convective term we choose
$${\bf F}\eql -{4acT^3\over 3\kappa \rho}\nabla T\ +\ \rho\left[T{dS\over d\log p} 
\right]^{3/2}{d\rbar\over d\phi}\nabla\phi\w,\eqno(122)$$
\ie 
$${\bf F}\eql \Chi(\phi)\nabla\phi\w,\w\Chi(\phi)\ \equiv\ -{4acT^3\over 3\kappa 
\rho}{dT\over d\phi}\ +\ \rho\left[T{dS\over d\log p} \right]^{3/2}{d\rbar\over 
d\phi}\w.\eqno(123)$$
Thus the equation of energy production and transport, taking into account the
meridional circulation current \vbf, is 
$$\nabla . \Chi \nabla\phi\  = \  \rho\epsilon - \rho T {\bf v}.\nabla S\w,
 \eqno(124)$$
where {\bf v} also satisfies the steady continuity equation
$$\nabla . \rho {\bf v}\  = \  0\ \  . \eqno(125)$$
Note that we are considering only the {\it rotational} component to the 
distortion of the star, which is approximately constant over many orbits, 
whereas the companion-induced distortion fluctuates on the timescale of one 
orbit, and produces the tidal velocity field of the previous Section.
\par We first establish that the 
circulation term carries no {\it net} energy across an equipotential surface,
\ie\  that $\int\rho T {\bf v}.\nabla S\ dV = 0$, where the integral is over 
the interior of an equipotential surface. From thermodynamics and hydrostatic
equilibrium,
$$TdS\  = \  dU+pd{1\over\rho}\  = \  d(U+{p\over\rho})-{1\over \rho}dp\  = 
d(U+{p\over\rho}+\phi)\ \  .\eqno(126)$$
Hence, using (125),
$$\int\rho T {\bf v}.\nabla S\ dV\  = \  \int \rho {\bf v}.\nabla(U+{p\over\rho}
+\phi)\th dV \  = \  \int (U+{p\over\rho}+\phi)\rho {\bf v}.d\vecSigma
\ \ .\eqno(127)$$
Since the expression in parentheses is constant on an equipotential it can come 
outside the last integral, and since $\int \rho {\bf v}.d\vecSigma=0$ by Eq. (125) we have the result. Hence
$$\int {\bf F}.d\vecSigma\eql \int \Chi\nabla \phi .d\vecSigma\eql \int\rho
\epsilon dV\eql L \ \  .\eqno(128)$$
Since $\Chi$ is constant on equipotentials, we can write this, using Eq. (116), as
$$L\  = \  \Chi \int\nabla\phi . d\vecSigma \  = \  \Chi K \ \  , \ \ \ie \ \ 
\Chi\  =\  L/K\ \  . \eqno(129)$$
\par Using Eqs (6), (115), (117) and (129), Eq. (124) becomes
$$ {L\over K}\th {dK\over dV}+\vert\nabla\phi\vert^2{d\over d\phi}\th 
{L\over K}\  = \  
{dL\over dV}-\rho T {dS\over d\phi}{\bf v}.\nabla\phi\ \  ,\eqno(130)$$
and so, after some manipulation,
$$\rho T{dS\over d\phi}\ v_{\perp}\eql \left({1\over \vert\nabla\phi\vert}
\int\vert \nabla\phi\vert\th d\Sigma\ -\ \vert\nabla\phi\vert\int{d\Sigma 
\over\vert\nabla
\phi\vert}\right)\ {d\over dV}\th{L\over 4\pi Gm -2\Omega^2 V}\ ,\eqno(131)$$
where $v_{\perp}$ is the component of \vbf\ in the direction of $\nabla\phi$.
The tangential component of \vbf\ then comes from continuity, Eq. (125).
\par The analysis of $\S$ 2, in particular Eq. (9), allows us to estimate 
the angular-dependent term in Eq. (131) for the circulation velocity driven
by the uniform rotation, in terms of the distortion parameter $\alpha$ and its
surface value (Equn 15a). It is
$$ {1\over \vert\nabla\phi\vert}
\int \vert\nabla\phi\vert\th d\Sigma\ -\ \vert\nabla\phi\vert\int{d\Sigma\over
\vert\nabla\phi\vert}\ \approx\ -{8\pi\Omega^2 r_1^3\over 3Gm_1(1-Q)}\th{r^2\over
\alpha_1}{d\th r\alpha\over dr}\th P_2(\cos\theta)\w.\eqno(132)$$
In addition to the rotationally-driven circulation of Equn (132), there
is in principle a circulation due to the part of the potential that comes from
$*2$'s gravity. But unless and until $*1$ is brought into synchronism, and the
orbit circularised, this contribution will fluctuate about zero with the 
period of $*1$'s relative rotation. It will therefore be insignificant until
synchronism is reached.
\par It is evident that the velocity field $v_{\perp}$ given by Eq. (131) is singular
on surfaces $dS/d\phi =0$ and $\rho=0$. For mathematical consistency we require
that there be a current sheet of zero thickness at each of these singular 
surfaces, to return the flow, which diverges out of the polar point on these 
surfaces from both sides (or else converges into it, also from both sides).
Presumably the flow in and near this sheet is turbulent because of the high 
shear, and this turbulence provides viscosity which renders the flow finite in 
practice.
\par There are many complexities in the study of meridional circulation
(Mestel 1965), including for example the influence of composition gradients,
and the redistribution of angular velocity from the (supposed) uniform state.
Nevertheless in the surface layers of rapidly rotating stars it could well be
the second most important source of turbulence, after convection. For recent work on meridional circulation, cf. the papers by Zahn (1992), Urpin et al. 
(1996), and Talon et al. (1997).
\section{Conclusions}
\par We have set out a formalism for developing the equations governing the
spin and orbital evolution of a binary subject to the `equilibrium tide' model
of tidal friction, along with the non-dissipative effect of the quadrupole
moment of the tide. We believe it may be helpful to have all the analysis in 
one place, and the analysis of the orbit by the use of the LRL vector 
seems considerably more efficient, as well as more transparent, than by other 
means. It leads directly
to formulae applicable to any degree of inclination of the star to the orbit,
and not just to small inclinations (Hut 1981). We also believe it may be 
helpful to have shown that the normal equation for tidal friction, based on the picture 
that the tidal bulge lags the line of centres by some small constant time,
follows quite directly from a more physical model in which the dissipation is
related to a positive-semidefinite function of the rate of change of the tidal
deformation (as measured by the quadrupole tensor) in the frame that rotates 
with the star. The effective lag time $\tau$ is 
related to the rate of dissipation $\sigma$ in our model by
$$\tau\eql {3\sigma r_1^5\over 4G}{Q\over 1-Q}\w .\eqno(133)$$
\par We have included a tentative determination of $\sigma$ (Eq. 113) in terms 
of a hypothetical isotropic viscosity due to the internal turbulent velocity 
distribution in the star. This is a much more uncertain  area, since the 
internal dynamics, particularly but not exclusively as it relates to 
turbulence, is far from clear, and very different formulations have been 
presented in the literature (Zahn 1978, Scharlemann 1982, Campbell \& 
Papaloizou 1983, Tassoul \& Tassoul 1990, Rieutord \& Zahn 1997). 
Regarding convectively driven turbulence, we note that our weight factor 
$\gamma$ in Eq. (113) is not especially small in the convective core of 
the $8\Msun$ star, being $\ts 0.003 - 0.01$ (Fig. 1a). This is larger than is
expected with the rather inaccurate approximations $\alpha\propto r^3, \beta
\propto r^4, \gamma \propto r^8$. We propose in the future to evaluate $\sigma$
from convection in a variety of stellar models, but a provisional estimate  
suggests a timescale $t_{\rm TF}\ts 30\th$yr, for $r_1\ts a, m_1\ts m_2 \ts 8\Msun$. 
\par Some further turbulence may be due to rotationally driven circulation 
($\S$ 7), which, while presumably not actually 
singular at the surface and at the boundaries of convection zones (of which 
there are often one or two near the surface even in stars with `radiative' 
envelopes), may be more significant there than is usually supposed. 
Another source might be differential rotation, itself brought about because
tidal friction is likely to operate most strongly on the outer layers.
Differential rotation leads to shear which should lead to turbulence.
\par It has long been suspected (Maeder 1975) that stars have larger convective 
cores than can be accounted for by `standard' assumptions about convective 
stability. Although several different models for this have been proposed, they 
are not consistent with each other (Eggleton 1983). But even in the absence of a
definitive model, one can compare theoretical models containing an artificial 
amount of `enhanced mixing' with those (relatively few) observed binaries, the 
$\zeta$ Aur binaries, which have (a) a highly evolved (red or orange supergiant)  component, and (b) fundamental data (masses {\it etc.}) accurate to better than 
5\%. Schr{\"o}der \etal (1997) show that $\zeta$ Aur stars give best agreement with theoretical models that mix cores of about 40\% more mass than `standard'
models at $\ts 8\Msun$. Thus there is almost certainly more motion taking place 
in stars than the simplest theoretical concepts allow, and this means that an 
estimate such as Eq (113) for $\sigma$ is bound to be a lower limit; by how much remains to be determined.
\par Although there is much disagreement in detail about the kinds of velocity 
fields to be expected in stars, including disagreement about the basic
assumption of hydrostatic equilibrium from which all the calculation of $\S$s
2--7 stem, we feel that at least this last assumption is reasonably secure at
the level which we require. Velocities would have to be sonic throughout the 
star for hydrostatic equilibrium to be seriously modified. Perhaps they may 
be this high very close to the photosphere, in very rapidly rotating stars 
and/or stars with surface convection, but hardly in deeper layers. Our 
Eq. (113) gives quite substantial dissipation from the deeper layers, 
including the core. Our determination of $\sigma$ in terms of $wl$ is 
tentative given that the distribution of turbulence is not well-known, but 
we believe our answer is `definitive' otherwise, particularly since 
it leads to just the same functional form for dissipation as does the earlier, 
more general, treatment of $\S$ 3 which makes no reference to a specific 
agent of dissipation.
\par We believe it may be better for the present to think of $\sigma$ as a 
parameter that might be amenable to observation. We consider briefly three
situations where $\sigma$, or at least limits to $\sigma$, might be measurable.
\halfline
\sep {\bf (i) PSR 0045-716}:
This radio pulsar in the SMC (51.2$\th$d, $e=0.808$; Kaspi \etal 1994) is 
unusual in that the star it orbits, an apparently normal B1V star, is not a Be 
star and therefore there is little or no accretion to perturb the orbit in an 
erratic fashion. The highly eccentric orbit is quite wide  so that even at 
periastron the stars are not very close: the separation between the two stars at periastron is approximately 4 times the radius of the companion B-star (Kaspi et al. 1996). The variation of the orbit with time is already measurable: the 
orbital period divided by the time derivative of the orbital period is $5\times 10^5\th$yr (Kaspi et al. 1996).  This variation may be due predominantly to 
tidal friction. In a reasonable interval of time rates of change of period, 
eccentricity and inclination should all be measurable. Our model gives equations which make a definite prediction about the time-variation, apart from the 
dissipation rate $\sigma$, and which may therefore be testable. This might not 
be quite definitive, since we need to know such things as both masses, and the 
orbital inclination, rather than just the (very well-determined) mass function; 
but in principle all the unknowns might be determined by fitting over a long 
stretch of observation.  Lai et al. (1995) have made a strong case that the spin
axis of the stars is inclined with respect to the total angular momentum of the system. The analysis of the dynamical tide for this system was carried out by Lai (1996), Kumar \& Quataert (1997), and in more detail by Kumar \& Quataert, 
(1998), and Lai (1998). In principle, all of $\edot,\etadot,\Chidot,\psidot$
might be measurable in time.
\halfline
\sep {\bf (ii) $\lambda$ Tau}: This is a well-known triple system in which 
the third body is very close to the 4d eclipsing pair ($P_{\rm out}\ts 33\th$d; 
Struve \& Ebbighausen 1956, Fekel \& Tomkin 1982). The orbits are quite probably parallel, or conceivably anti-parallel (S{\"o}derhjelm 1975). In the latter
case the total angular momentum would be close to zero. The third body should 
be injecting a fluctuating eccentricity into the close pair, on a timescale 
of days. For point masses,  this fluctuation would be of order 0.007, if the 
outer orbit is circular. Such a 
fluctuation in the eccentricity of a semidetached pair might be expected to 
have a dramatic effect on the rate of mass transfer, since the pressure 
scale-height in the atmosphere, relative to the radius, is probably at 
least ten times smaller. The absence of such an effect suggests
that tidal friction is quite efficient at removing eccentricity; but too-strong
tidal friction could mean that the inner orbit shrinks rapidly compared with
a reasonable estimate of its age on the grounds of nuclear evolution. Thus we
might expect to find both a lower and an upper limit to the dissipation
constant $\sigma$ from this system.
\halfline
\sep {\bf (iii) Algol}: The well-known triple star Algol (Tomkin \& Lambert 
1978, Hill \etal 1971, Labeyrie \etal 1974, Popper 1980) has a long-period 
orbit ($P_{\rm out}=1.86\th$y) which, from VLBI measurements, is inclined at 
$\ts 100\arcdeg$ to the orbit of the inner eclipsing pair (Lestrade \etal 1993). 
In such a highly-inclined system, if the stars are treated as three point 
masses the inner orbit should cycle in eccentricity between $e\ts 0$ and 
$e\ts 0.98$, on
a timescale of $\ts P_{\rm out}^2/P_{\rm in}\ts 500\th$y, while preserving
both inner and outer periods. It presumably does not do so, and this might
partly be because of tidal friction. However, the quadrupolar distortion of 
the stars even in the absence of dissipation should lead to apsidal motion 
in the close pair, which would be quite rapid compared to the apsidal motion 
driven by the third body. Since the large cycles in eccentricity depend
rather critically on the rate of apsidal motion driven by the third star,
the extra apsidal motion of the quadrupolar distortion may prevent these
cycles from being large; but we might still expect larger fluctuations
on account of the inclination than we would if the orbits were parallel. 
It is not clear that there is a lower limit to $\sigma$ from the (presumed) 
absence of large cycles in eccentricity. But at the same time tidal friction 
must not transfer angular momentum from the inner pair to the outer pair so fast that the inner pair would have difficulty in lasting for a nuclear timescale, 
since the system has in fact existed long enough for considerable nuclear evolution. Thus there may still be an upper limit. Other close triples such as 
DM Per (Hilditch \etal 1986), VV Ori (Duerbeck 1975), HD109648 (Jha \etal 1997) 
and $\zeta^2$ CrB (Gordon \& Mulliss 1997), with {\it outer} periods of 100d, 
119d, 121d and 250d, may also give upper limits.
\halfline
\par The first of the above three situations gives a reasonable chance of either
confirming or ruling out the equilibrium tide model; all the initial conditions
as well as $\sigma$ will have to be treated as free parameters, which might be
determined by fitting over a considerable stretch of time. An analysis of the
two triple systems ($\lambda$~Tau, $\beta$ Per) is currently being undertaken.
\acknowledgments
\par LGK and PPE are grateful to NATO for financial support with grant
CRG 941288. PPE is grateful to Dr. D. C. Heggie for recommending the LRL
formalism.   PH thanks Pawan Kumar for his comments on the manuscript.
The authors thank the anonymous referee for providing extra references and 
suggestions for adding appropriate disclaimers.

\newpage

\end{document}